\documentclass[12pt,a4paper,oneside,onecolumn]{article}
\usepackage[]{fontenc}

\begin{document}

\begin{center} 

{\Large \bf The Chandrasekhar limit for quark stars} \\

\vspace{0.5in} 

{\bf Shibaji Banerjee{\footnote{email: phys@bosemain.boseinst.ernet.in}},
Sanjay K. Ghosh{\footnote{email: phys@bosemain.boseinst.ernet.in}} and
Sibaji Raha{\footnote{email: sibaji@bosemain.boseinst.ernet.in}}} 

Department of Physics, Bose Institute, \\ 93/1, A. P. C. Road, Calcutta
 700 009, INDIA 

\end{center} 

\begin{abstract} The Chandrasekhar limit for quark stars is
evaluated from simple energy balance relations, as proposed by Landau for
white dwarfs or neutron stars. It has been found that the limit for quark
stars depends on, in addition to the fundamental constants, the Bag
constant.

\end{abstract}

Compact objects like the white dwarfs and neutron stars have been
a topic of interest for several decades. These objects are produced
as the end product of the stellar evolution, \em {i.e.} \em , when the
nuclear
fuel of the normal stars have been consumed. White dwarfs are supported
against gravitational collapse by the degeneracy pressure of electrons
whereas for neutron stars, this pressure comes mainly from the degenerate
neutrons. It is well known that both these classes of collapsed stars have
a maximum mass ( \( \sim 1.5 M_\odot \) ),
beyond which they collapse to black holes \cite{chandra,landau,shapiro}.
Although the Chandrasekhar limit refers strictly to white dwarfs, the
limiting mass for neutron stars is also loosely called the Chandrasekhar
limit, primarily because the limits in the two cases turn out to be the
same
\cite{landau}.

The underlying quark structure of the nucleons suggests a possible
hadron - quark phase transition at high density and/or temperature.
This, in turn, implies the possible existence of another kind
of compact objects, namely quark stars \cite{baym,alcock,qsrev}.
The suggestion of Witten\cite{witten} that the strange quark matter
may be the true ground state of the strongly interacting matter 
started a new era in this subject. In this circumstance, quark stars, if
they are formed, would preferably convert to strange stars, comprising
{\it u, d} and {\it s} quarks, under weak interaction. Several authors
( for example, \cite{jpg,ghosh1} ) have used different models to
understand the properties of strange stars. For a review, see
\cite{qsrev}. 

The unique feature of strange stars, which separates them from
other compact objects, is that these need not be the direct product of
stellar evolution. These are most likely to be produced due to the hadron
quark phase transition inside the neutron stars as the density is
high enough to favor such a transition. For such quark stars, the maximum
mass would indeed be almost the same as that for neutron stars. On the
other
hand, it is also conceivable that if a large amount of quark matter exists
in the universe as a relic of the cosmological quark-hadron phase
transition
\cite{apj}, it could clump under gravitational interaction and even form
invisible quark galaxies \cite{sweden}. The `stars' of such a galaxy would
be strange stars which do not evolve from neutron stars and thus are not
governed by the Chandrasekhar limit for neutron stars. It is therefore
most
pertient to ask if there exists, just like the case of ordinary compact
stars,
an upper limit on the strange stars beyond which they would be
gravitationally unstable. Starting with the seminal work of Witten
\cite{witten}, most authors have concentrated on solving the
Tolman-Oppenheimer-Volkov (TOV) equation (see, for example,
\cite{shapiro})
numerically for the quark matter equation of state. While
the results show that indeed there exists a limiting mass for quark stars
(which is very close to that for neutron stars), there is no \em a priori
\em argument to prove that such a limit should exist or that it should
depend mostly on fundamental constants, as is the case for the ordinary
compact stars \cite{shapiro}. Our aim in this letter
is to show analytically, \em from first principles \em , that such a limit
exists for compact quark stars and that it is indeed determined mostly by
universal constants. 

We start with the most general and simple picture of energy balance,
proposed
by Landau \cite{landau}. The strange star is composed of a 3 flavour
system
of \em massless \em quarks, confined in a large bag \cite{baym}
characterized
by a constant energy density \( B \). As in the case of white dwarfs and
neutron stars, the equilibrium should occur at a minimum of the total
energy
per fermion \( e \), where \( e \equiv e_{F}+e_{G} \), in which \( e_{F}
\)
is the fermi energy  and \( e_{G} \) is the gravitational energy per
fermion.
The crucial difference between the cases of quark stars and the ordinary
compact stars lies in the estimation of the (Newtonian) gravitational
energy,
considered to be a macroscopic quantity. For ordinary compact stars, this
mass is due almost entirely to the baryons. For quark stars, however, the
total mass is the total ( thermodynamic as well as the confining ) energy
in
the star. For the purpose of estimating the gravitational energy per
fermion,
one can then define an effective quark mass which incorporates both these
contributions. A suitable prescription for such a definition was
formulated
quite some time ago \cite{dden}.

The number density of fermions is related to the chemical potential as
\[ n=\frac{g}{6\pi ^{2}}\mu ^{3}\]
which dictates that 

\begin{equation}
\label{mu-n}
\mu =(\frac{9\pi }{2g})^{\frac{1}{3}}\frac{N^{\frac{1}{3}}}{R}
\end{equation}

In the above relations, \( n \) is the number density, \( N \) the total
number of fermions in a star of radius \( R \), \( g \) the statistical
degeneracy factor and \( \mu  \) is the chemical potential.

The fermion energy density is given by

\begin{equation}
\label{eden}
\varepsilon _{F}=\frac{g}{8\pi ^{2}}\mu ^{4}
\end{equation}

and hence the fermi energy per particle of the quarks becomes 

\begin{equation}
\label{ef}
e_{F}=\frac{\varepsilon _{F}}{n}
=\frac{3}{4}(\frac{9\pi }{2g})^{\frac{1}{3}}\frac{N^{\frac{1}{3}}}{R}
\end{equation}

The mass \( M \) of the star can be written in terms of \( N \) and
\( B \) (the bag constant), if the density
\( \rho (r) \) in the star is assumed to be roughly constant throughout
the volume of the star. Hence using eq.(\ref{ef}),

\begin{equation}
\label{Mass1}
M=\int _{0}^{R}4\pi r^{2}\rho (r)\, dr
=\frac{4}{3}\pi R^{3} B + e_{F} N
=\frac{3}{4}(\frac{9\pi }{2g})^{\frac{1}{3}}\frac{N^{\frac{4}{3}}}{R}
+\frac{4}{3}\pi BR^{3}
\end{equation}

Extremising the mass \( M \) (eq. \ref{Mass1}) with respect to \( R \)
gives,

\begin{equation}
\label{Ef=3b}
(\frac{9\pi }{2g})^{\frac{1}{3}}\frac{N^{\frac{4}{3}}}{R^4}
=\frac{16}{3}\pi B     \nonumber \\
\Rightarrow \\
\varepsilon _{F}=3 B
\end{equation}

Substituting eq.(\ref{Ef=3b}) in the expression for \( M \)
(eq. \ref{Mass1}),
\begin{equation}
\label{Mb}
M = 4 B V = \frac{16}{3} \pi B R^3
\end{equation}

Note that this is very similar to the condition obtained for hadronic bags
\cite{mit}. The task at hand then is to find the \( R \) for which the
total energy per fermion would be maximum.

The gravitational energy per fermion \( e_{G} \) is

\begin{equation}
\label{eg1}
e_{G}=-\frac{GM{m_{eff}}}{R}
\end{equation}
where \( m_{eff} \) is the effective quark mass inside the star. Assuming
that the effective quark mass contributes to the total star mass \( M \),
one can write for a strange star with \( N \) quarks,
\begin{equation}
\label{Mm}
M = N m_{eff} \Rightarrow m_{eff} = \frac{4 B}{n}
\end{equation}

As mentioned above, the effect of confinement in a quark matter system was
shown \cite{dden} to be incorporable  in the effective quark mass, which,
the
quarks being fermions, coincides with the quark chemical potential . As a
result, one gets, in the limit of vanishing quark density \cite{dden},
\[ \mu = \frac{B}{n} \] .
This, together with the eq.(\ref{Mm}), gives
\begin{equation}
\label{qmas}
m_{eff} = 4 \mu
\end{equation}
where all the energy ( thermodynamic and confining ) is included in the
effective gravitational mass of the quarks inside the strange star.

Using equations (\ref{eg1}), (\ref{Mm}) and (\ref{qmas}) we get 

\begin{equation}
\label{eg2}
e_{G}=-\frac{64}{3}(\frac{9\pi }{2g})^{\frac{1}{3}}G \pi B R
{N^{\frac{1}{3}}}
\end{equation}

Minimising the total energy \( e=e_{F}+e_{G} \) with respect to \( N \),
we
get the expression for maximum value of \( R \) as

\begin{equation}
\label{rmax}
R_{max}=\frac{3}{16}\frac{1}{\sqrt{\pi GB}}
\end{equation}

Finally, the maximum mass of the strange star is computed by substituting
the
value of \( R_{max} \) (from equation \ref{rmax}) in equation ( \ref{Mb}).

\begin{equation}
\label{Mmax}
M_{max}=\frac{16}{3}{\pi B {R_{max}}^3}
\end{equation}

The chemical potential \( \mu \) can be evaluated in terms of \( B \)
using equations (\ref{eden}) and (\ref{Ef=3b}). Substituting this in
eq.(\ref{Mm}) gives the value of \( N_{max} \).
The values of \( R_{max} \),
\( M_{max} \) and \( N_{max} \) are tabulated below for various values of
the Bag constant \( B \).

\vspace{0.30cm}
{\centering \begin{tabular}{|c|c|c|c|}
\hline 
\( B^{1/4}\, (MeV) \)&\( R_{max}\, (Km) \)&\( \frac{M_{max}}{M_{\odot }}
\)&\( N_{max} \) \\
\hline 
\hline 
145&12.11&1.54&1.55 \( \times \) 10$^{57}$\\
\hline 
200&6.36&0.81&5.90 \( \times \) 10$^{56}$\\
\hline 
245&4.24&0.54&3.21 \( \times \) 10$^{56}$\\
\hline 
\end{tabular}\par}
\vspace{0.30cm}

We thus find that even for quark stars, there does exist a limiting mass,
the so-called Chandrasekhar limit, which is mostly determined by the
universal constants ( \( G \) as well as \( \hbar \) and \( c \), which do
not
occur explicitly due to our use of the naturalised units ) and the Bag
energy \( B \). Although treated as a parameter here,
the Bag energy too has, roughly, the status of another universal constant,
being the difference between the non-perturbative and the perturbative
vacua
of Quantum Chromodynamics. The physical radius \( R_{max} \),
corresponding
to the maximum mass as well as the maximum mass itself, are independent of
the number of flavours, as seen from equations (\ref{rmax}, \ref{Mmax}).
Although \( N_{max} \) depends on the statistical degeneracy factor \( g
\)
( or equivalently, the number of flavours), the dependence is extremely
weak,
as can be readily checked from equation (\ref{mu-n}). In fact, we have
verified that there is almost no difference in \( N_{max} \) between the
cases with \( g \) = 2 and 3. This, in turn, implies that the assumption
of
massless quarks ( even for \( s \) quarks ) does not affect these results.

Admittedly, our purpose here has been mostly illustrative. Nevertheless,
these limits agree very well with those found in the numerical solutions
of
the TOV equation (see, for example,
\cite{witten} ). Even under the simplifying assumption of a constant
density
profile, necessary for an analytical solution, the scaling behavior
( \( R_{max} \propto B^{-1/2}, \, M_{max} \propto B^{-1/2} \) , obtained
from the numerical solutions ) hold true. This proves that the simple
picture
presented here adequately incorporates the essential physics.

To conclude, we have shown that there exists a limiting mass (the
Chandrasekhar limit) for compact quark stars, beyond which they would be
gravitationally unstable. As with white dwarfs and neutron stars, this
mass
depends mostly on universal constants.

The work of SB and SKG were supported in part by Council of Scientific and
Industrial Research, Govt. of India.


\begin{thebibliography}{25}
\bibitem{chandra} S. Chandrasekhar, {\it An Introduction to the Study of
Stellar Structure}, Dover Publications, New York, 1958 and references
therein.
\bibitem{landau} L. D. Landau, {\it Phys. Z. Sowjetunion} {\bf 1}, 285
(1932), reprinted in {\it Collected Works of L. D. Landau} (Ed. D Ter
Haar),
p. 60, Pergamon Press, Oxford, 1965.
\bibitem{shapiro} S. L. Shapiro and S. A. Teukolsky, {\it Black Holes,
White
Dwarfs and Neutron Stars}, John Wiley \& Sons, New York, 1983.
\bibitem{baym} G. Baym and S. -A. Chin, {\it Phys. Lett.} {\bf B62}, 241
(1976)
\bibitem{alcock} C. Alcock, E. Farhi and A. Olinto, {\it Ap. J.}
{\bf 310}, 261 (1986)
\bibitem{qsrev} See, for example, J. Madsen, in : {\it Hadrons in dense
matter and hadrosynthesis}, Lecture Notes in Physics, Springer Verlag,
Heidelberg (to appear); astro-ph/9809032 (1998)
\bibitem{witten} E. Witten, {\it Phys. Rev.} {\bf D30}, 274 (1984)
\bibitem{jpg} P. Haensel, J. L. Zdunik and R. Schaeffer, {\it Astron.
Astrophys.} {\bf 160}, 121 (1986)
\bibitem{ghosh1} S. K. Ghosh and P. K. Sahu, {\it Int. J. Mod. Phys.}
{\bf E2}, 575 (1993)
\bibitem{apj} J. Alam, S. Raha and B. Sinha, {\it Ap. J.}  {\bf 513},
572 (1999)
\bibitem{sweden} D. Enstrom {\it et al}, astro-ph/9802236 (1998); S.
Fredriksson {\it et al}, astro-ph/9810389 (1998)
\bibitem{dden}G. N. Fowler, S. Raha and R. M. Weiner, {\it Z. Phys} {\bf
C9},
271 (1981); M. Pl\"{u}mer, {\it Quark - Gluon - Plasma und
vielfacherzeugung
in der Starken Wechselwirkung}, Doctoral dissertation, Philipps
Universit\"at,
Marburg/Lahn, Germany (1984).
\bibitem{mit} A. Chodos {\it et al}, {\it Phys. Rev.} {\bf D9}, 3471
(1974) 
\end{thebibliography}
\end{document}